\documentstyle[prl,aps]{revtex}
\begin{document}
\twocolumn[\hsize\textwidth\columnwidth\hsize\csname
@twocolumnfalse\endcsname

\title{On the Thomas-Fermi approximation
in the bulk of trapped Bose-Einstein condensed gases}
\author{Gyula Bene$^1$ and P\'eter Sz\'epfalusy$^2$}
\address{$^1$Institute for Theoretical Physics, E\"otv\"os University,
     P\'azm\'any P\'eter s\'et\'any 1/A, H-1117 Budapest, Hungary\\
$^2$Department of Physics of Complex Systems, E\"otv\"os University,
     P\'azm\'any P\'eter s\'et\'any 1/A, H-1117 Budapest, Hungary
and Research Institute for Solid State Physics and Optics of the Hungarian Academy
of Sciences, P.O.Box 49, H-1525 Budapest, Hungary }
\date{\today}
\maketitle

\begin{abstract}
Corrections to the Thomas-Fermi-type solution of the 
Gross-Pitaevskii equation are inevitable in order to get correctly
the frequencies of the low lying modes out of the Bogolyubov equations.
These corrections are important in the bulk, too, thus the failure
of the Thomas-Fermi approximation is not confined to the surface.
We discuss this effect quantitatively and consider similar phenomena
of spin fluctuations in Bose-Einstein condensed gases in an optical trap. 
\end{abstract}

\vskip2pc]
\narrowtext

Several properties of the recently much studied trapped Bose-Einstein condensed
alkali metal vapors can be reasonably well explained in terms of the 
Gross-Pitaevskii equation and of the Hartree-Fock-Bogolyubov-Popov equations
(for a review see \cite{surface}).
These equations yield good approximations for the condensate shape and size, 
the
chemical potential, the critical temperature and for the excitation 
frequencies.
Note that they do not describe the damping of the collective excitations, 
as the
Bogolyubov approximation neglects terms of the Hamiltonian that are 
of third and fourth
order in the field operator (after having used the usual Bogolyubov 
prescription).
Although themselves being 
approximations (even within the framework of nonrelativistic
many body theory), they are usually subjects 
of further approximations to get solutions. This
is not as if a numerical approach were so difficult, but rather to get a deeper
understanding of the underlying physics and to clarify the significance of the 
involved mechanisms. The quantum hydrodynamical description introduced
by Stringari \cite{hydro1} for trapped Bose gases has proven to be important
in case of large particle numbers.

At large condensates, and for repulsive interaction
(i.e., for $g>0$, as in case of rubidium and sodium vapours) 
one can apply the Thomas-Fermi approximation for the solution of 
the Gross-Pitaevskii
equation (\ref{GP})
\begin{eqnarray}
\left[\hat {\cal L} - g\left|\Phi_0(\vec{r})\right|^2 \right]\Phi_0(\vec{r})=0
\label{GP}
\end{eqnarray}   
where
\begin{eqnarray}
\hat {\cal L}=-\frac{\hbar^2}{2m}\triangle + V(\vec{r})-\mu + 
2g\left|\Phi_0(\vec{r})\right|^2+2g
\langle\hat \psi^\dagger(\vec{r})\hat \psi(\vec{r})\rangle
\label{L}
\end{eqnarray}   

In the above equation $\Phi_0(\vec{r})$ stands for the 'condensate 
wavefunction',  $V(r)$ denotes the trap potential, the average in the
last term on then right hand side gives the density of the noncondensed
atoms
and $\mu$ is the chemical
potential which is connected with the particle number and the temperature 
through
$
N=\int d^3\vec{r}\left[\left|\Phi_0(\vec{r})\right|^2+
\langle\hat \psi^\dagger(\vec{r})\hat \psi(\vec{r})\rangle\right]$.
In the following we shall consider such situations only where 
$\Phi_0(\vec{r})$ can be taken as real. The Thomas-Fermi approximation
 amounts to the omission of the kinetic term 
$-\frac{\hbar^2}{2m}\triangle\Phi_0(\vec{r})$. Thus one gets the equation
\begin{eqnarray}
 g\left(\Phi_0^{TF}(\vec{r})\right)^2+2g
\langle\hat \psi^\dagger(\vec{r})\hat \psi(\vec{r})\rangle
=\mu - V(\vec{r})
\label{TF}
\end{eqnarray}   
Certainly, this equation cannot hold when $V(\vec{r})>\mu$, 
as the left hand side
of Eq.(\ref{TF}) cannot be negative. Actually it is well known that 
in the regime
$V(\vec{r})\lesssim \mu$ which defines approximately the surface 
of the condensate 
the Thomas-Fermi approximation breaks down, which  becomes
obvious if one inserts $\Phi_0^{TF}(\vec{r})$ into the neglected kinetic term.
For a spherical condensate this surface layer is of width 
$\delta \approx \frac{1}{4}
\left(\frac{\hbar\omega_0}{\mu}\right)^{2/3}r_{TF}$ 
where $r_{TF}=\sqrt{\frac{2\mu}{m\omega_0^2}}$
stands for the Thomas-Fermi radius of the condensate and 
$\omega_0$ is the trapping frequency. Note that 
$r_{TF}\propto N^{1/5}$ and $\delta\propto N^{-1/15}\propto N^{-4/15}r_{TF}$.
Surface corrections to the ground state energy have been 
thoroughly discussed and calculated
\cite{surface}, \cite{Ruprecht}, \cite{Dalfovo}. 
From the point of view of the chemical potential and the condensate
shape corrections to the Thomas-Fermi approximation are important 
only near the surface 
of the condensate. In contrast,  
from the point of view of the excitation frequencies
corrections to the Thomas-Fermi approximation are inevitable 
in the bulk as well \cite{hydro3},\cite{fetter-rokhsar}.
First of all, recall that at low temperatures the lower excitation frequencies
the Hartree-Fock Bogolyubov-Popov equations

\begin{eqnarray}
\hat {\cal L} u_j(\vec{r})-g\left|\Phi_0(\vec{r})\right|^2v_j(\vec{r})=
E_j u_j(\vec{r})\nonumber\\
\hat {\cal L} v_j(\vec{r})-g\left|\Phi_0(\vec{r})\right|^2u_j(\vec{r})=
-E_j v_j(\vec{r})
\label{Bog}
\end{eqnarray}
are rather well approximated by the solution of the equation
\begin{eqnarray}
-\frac{g}{m}\nabla \left(\left(\Phi_0\right)^2\nabla 
\varphi\right)=\omega^2 \varphi\label{hyd}
\end{eqnarray} 
called hydrodynamic approximation.
For cylindrically symmetric harmonic potential 
(which is relevant for the actual experimental
situation) and by using the Thomas-Fermi expression (\ref{TF}) at zero
temperature, Eq.(\ref{hyd}) has been solved analytically\cite{hydro1},
\cite{hydro2},\cite{hydro3} and the resulting $\omega$
values agree reasonably with low temperature experimental data.
Note that at zero temperature 
Eq.(\ref{GP}) is uncoupled from Eq.(\ref{Bog}).

When the radial trapping frequency $\omega_r$ is smaller than the
axial one $\omega_z$ (disc-shaped condensate), the lowest nonzero eigenvalue is
$\omega=\omega_r$.

Nevertheless, if
one inserts the Thomas-Fermi expression for $\Phi_0$ into the Hartree-Fock-Bogolyubov-Popov
equation (\ref{Bog}), one obtains completely erroneous results, as we now explicitly show. 
Indeed, Eqs.(\ref{Bog}), (\ref{TF}) imply
\begin{eqnarray}
-\frac{\hbar^2}{2m}\triangle G_j=E_j F_j\label{er1}\\
-\frac{\hbar^2}{2m}\triangle F_j + 2g\left(\Phi_0^{TF}\right)^2 F_j=E_j G_j\label{er2}
\end{eqnarray}
where $G_j=u_j+v_j$ and $F_j=u_j-v_j$. 
At low energies, i.e., when $\hbar \tilde \omega \ll \mu \sim g\left(\Phi_0^{TF}(0)\right)^2$ 
the first term on the left hand side of Eq.(\ref{er2}) is much smaller
than the second term, thus can be neglected. Therefore, instead of (\ref{er2})
\begin{eqnarray}
2g\left(\Phi_0^{TF}\right)^2 F_j=E_j G_j
\label{er2b}
\end{eqnarray}
may be used.  Eliminating $G_j$ 
one gets
$-\frac{\hbar^2 g}{m}\triangle \left(\Phi_0^{TF}\right)^2 F_j=E_j^2F_j\,$.
Introducing 
$\tilde \varphi=\Phi_0^{TF} F_j$
and 
$\tilde \omega=\frac{E_j}{\hbar}$
one finally arrives at
\begin{eqnarray}
-\frac{g}{m}\Phi_0^{TF} \triangle\Phi_0^{TF} \tilde \varphi=\tilde
\omega^2 \tilde \varphi\,.\label{er4}
\end{eqnarray}
This equation can be solved with the same techniques as Eq.(\ref{hyd}), i.e., 
its validity is extended up to the surface of the condensate 
(as defined by the vanishing of $\Phi_0^{TF}$) and then it is required that 
the solution does not diverge at the surface. Note that this induces the 
vanishing of the eigenfunction at the border, unlike in case of the hydrodynamical equation 
where the solution remains there nonzero.
The first 
eigenvalue at zero temperature is 
$
\tilde \omega=\sqrt{\omega_z^2+2\omega_r^2}$. One may prove the inequality
$-\int d^3 \vec{r} \tilde \varphi \Phi_0^{TF} \triangle \left(\tilde \varphi \Phi_0^{TF} \right)\ge
-\int d^3 \vec{r} \tilde \varphi^2 \triangle \left(\Phi_0^{TF} \right)^2$ which readily implies
that $\tilde \omega^2\ge \omega_z^2+2\omega_r^2$. Equality holds for the eigenfunction $\tilde \varphi=\frac{1}{\sqrt{N}}
\Phi_0^{TF}$. 
Comparing $\tilde \omega$ with the experimentally confirmed result $\omega=\omega_r$ the obvious conclusion is that 
Eq.(\ref{er4}) is not correct. We emphasize that the source of the error is the
Thomas-Fermi approximation. Indeed, in Ref.\cite{fetter-rokhsar} Eq.(\ref{hyd}) 
has been derived from Eqs.(\ref{Bog}) by using the full Gross-Pitaevskii equation (\ref{GP}). 
That derivation has been done for the zero temperature case and for low excitation energies.
Taking into account the next correction to the Thomas-Fermi approximation,
one can derive the correct equation (\ref{hyd}) even at finite temperatures. 
To this end, let us write 
$
\Phi_0=\Phi_0^{TF}+\Phi_0^{(1)}$
where $\Phi_0^{TF}$ satisfies Eq.(\ref{TF}) and $\Phi_0^{(1)}$ is the correction term.
We determine the latter perturbatively (assuming that we are
sufficiently far from
the surface), i.e., we insert the Ansatz above into
Eq.(\ref{GP}) and keep the kinetic term $-\hbar^2/2m \triangle \Phi_0^{TF}$ while
neglect $-\hbar^2/2m \triangle \Phi_0^{(1)}$ and $\left(\Phi_0^{(1)}\right)^2$. 
We get
$
\Phi_0^{(1)}=\frac{\hbar^2}{4mg}\frac{\triangle \Phi_0^{TF}}{\left(\Phi_0^{TF}\right)^2}\,$.
Inserting these expresssions into (\ref{Bog}) we get instead of Eqs.(\ref{er1}), (\ref{er2})
\begin{eqnarray}
\left[-\frac{\hbar^2}{2m}\triangle + \frac{\hbar^2}{2m}\frac{\triangle \Phi_0^{TF}}{\Phi_0^{TF}}\right]G_j=E_j F_j\label{ok1}\\
\left[ -\frac{\hbar^2}{2m}\triangle + 2g\left(\Phi_0^{TF}\right)^2+ \frac{3\hbar^2}{2m}\frac{\triangle \Phi_0^{TF}}{\Phi_0^{TF}}\right]F_j=E_j G_j\label{ok2}
\end{eqnarray}
In Eq.(\ref{ok2}) we apply the same approximation 
as in case of Eq.(\ref{er2}), namely, we keep only the dominant term
$2g\left(\Phi_0^{TF}\right)^2 F_j$ on the left hand side,  
thus we arrive at Eq.(\ref{er2b}) again. Comparing now Eq.(\ref{ok1}) with
Eq.(\ref{er1}) it is clearly seen that the additional term on the left hand 
side cannot be neglected. In fact, for low excitation energies
the wavelength of the excitations is comparable to the condensate size,
thus the kinetic energy term is of the same order of magnitude
as the additional term on the left hand side of Eq.(\ref{ok1})\cite{hydro3}, as we show this below explicitly for $T=0$.
Let us emphasize that this is true even deeply in the bulk, notwithstanding
how small $\Phi_0^{(1)}$ is compared to $\Phi_0^{TF}$.
Note that in case of the Gross-Pitaevskii equation 
the correction term $\Phi^{(1)}_0$
should be compared to $\Phi_0^{TF}$, and they become of the same order
only in the surface region of the condensate. Unlike this, when
adding the Bogolyubov equations to get Eq.(\ref{ok1}), the lowest order
terms  
 (i.e., 
terms like $2g\left(\Phi_0^{TF}\right)^2 G_j$) cancel exactly, and therefore
the correction term $\frac{\hbar^2}{2m}\frac{\triangle \Phi_0^{
TF}}{\Phi_0^{TF}}G_j$ should be compared to the kinetic term 
$\frac{\hbar^2}{2m}\triangle G_j$. 
Eqs.(\ref{ok1}) and (\ref{er2b}) lead readily to
Eq.(\ref{hyd}). Thus taking into account the first correction to the
Thomas-Fermi approximation suffices to recover the hydrodynamic
approximation, while without it one gets the wrong result.
We want to investigate this question quantitatively.
Having solved Eq.(\ref{hyd}) at $T=0$ and using Eq.(\ref{er2b}) 
one can calculate the terms on the left hand side of Eq.(\ref{ok1})
separately. In case of $\omega_z>\omega_r$ (disc-shaped condensate)
we get for the ratio
$\frac{\hbar^2}{2m}\triangle G_j\left/\frac{\hbar^2}{2m}\frac{\triangle \Phi_0^{TF}}{\Phi_0^{TF}}G_j\right.$
the expression
\begin{eqnarray}
1+\frac{2\omega^2}{\omega_r^2}\left[
1+\frac{1}{1-\epsilon^2(\xi^2+1)}+\frac{1}{1-\epsilon^2(1-\eta^2)}\right]^{-1} 
\end{eqnarray}
where $\epsilon^2=1-\omega_r^2/\omega_z^2$ and $\xi$, $\eta$ are
the oblate spheroidal coordinates defined as
\cite{hydro2} $\rho=\sqrt{x^2+y^2}=\sqrt{\frac{2\mu}{m\omega_r^2}}\epsilon\sqrt{(\xi^2+1)(1-\eta^2)}$, 
$z=\sqrt{\frac{2\mu}{m\omega_r^2}}\epsilon \xi \eta$. 
For the lowest excitation energies $\omega$ is 
of the same order of magnitude as $\omega_r$, therefore
within the condensate the ratio of the two terms on the left hand side 
of Eq.(\ref{ok1}) is of order unity.

As a further demonstration that the correction term describes an important bulk
effect we exclude the surface region, hence we consider a situation when Eq.(\ref{er4}) holds
for $\xi<\xi_1<\xi_0=\sqrt{1/\epsilon^2-1}$, i.e., inside of the condensate,
while Eq.(\ref{hyd}) holds for $\xi_1<\xi<\xi_0$,
i.e., near the surface. Note that the Thomas-Fermi surface is
the ellipsoid $\xi=\xi_0$ and $\xi=\xi_1$ defines the
ellipsoidal bordering surface between the regions of validity
of Eqs.(\ref{er4}), (\ref{hyd}). 
Then the correction to the eigenvalues 
of (\ref{hyd}) comes from the bulk by construction. We calculate this
correction by perturbation theory in first order, i.e., we consider
Eq.(\ref{hyd}) as the unperturbed one and the additional term 
\begin{eqnarray}
-\frac{g}{m}\varphi \Phi_0^{TF}\triangle\Phi_0^{TF}
\end{eqnarray}
as a perturbation for $\xi<\xi_1$. The correction to the eigenvalues then
has the form
\begin{eqnarray}
\Delta \omega^2=-\frac{g}{m}
\int_0^{\xi_1} d\xi \int_{-1}^1 d\eta (\xi^2+\eta^2)\varphi^2 \Phi_0^{TF}\triangle\Phi_0^{TF}
\end{eqnarray}
provided that the unperturbed eigenfunction $\varphi$ is normed by
\begin{eqnarray}
\int_0^{\xi_0} d\xi \int_{-1}^1 d\eta (\xi^2+\eta^2)\varphi^2 
=1
\end{eqnarray}
In case of the lowest excitation frequency  one gets 
\begin{eqnarray}
\Delta\omega^2=\frac{\omega_r^2}{12\sqrt{1-\epsilon^2}}\Biggl\{
3\epsilon\xi_1 (\xi_1^2+1)\left[(5-4\epsilon^2)
\right.\Biggr.\nonumber\\\Biggl.\left.
-\epsilon^2(3-2\epsilon^2)(\xi_1^2+1)\right]
+6\frac{5-4\epsilon^2}{\sqrt{1-\epsilon^2}}{\rm arctanh}\left(\frac{\xi_1}{\xi_0}
\right)\Biggr.\\\Biggl.
-3\xi_1\frac{(5-4\epsilon^2)(\xi_1^2+3)-3\epsilon^2(\xi_1^2+1)^2}{\sqrt{1-\epsilon^2}}\arctan\left(\frac{1}{\xi_0}\right)
\Biggr\}\nonumber
\end{eqnarray} 

Certainly, the validity of this expression 
is confined for small $\xi_1$ values only (i.e., perturbation deeply in the bulk). 
In this regime we get in case of $\omega_r^2/\omega_z^2=1/8$ that
$\sqrt{\omega_r^2+\Delta\omega^2}-\omega_r\approx 6.74\omega_r \xi_1/\xi_0$.
The linearity in $\xi_1$ means a proportionality to the volume.
Note that the considered mode 
is just the Kohn mode whose frequency is known to be exactly the 
trapping frequency $\omega_r$. 

The effect due to the correction term 
is a consequence of inhomogenity (i.e., spatial
dependence of $\Phi_0^{TF}$) which is induced by 
the presence of the trapping potential. 
In the Thomas-Fermi approximation the 
effect depends only on the external potential and the number of particles.
Note that in case of a system contained in a box (with zero external potential inside)
the Thomas-Fermi solution is uniform and the correction
of the type discussed above is zero.
In the extreme situation of a single particle, however,
spatial dependence of the wave function can be a consequence
of the boundary conditions, as it is in case of a particle in a box.
Therefore, it is instructive to consider such a surface-induced
inhomogenity for a large number of particles. 
In case of a spherical box (with zero potential inside) we find
that the boundary induces a correction 
to the uniform Thomas-Fermi wave function which is comparable with it
near the surface and decays as $\exp(-k d)$ (where $k=2\sqrt{m\mu}/\hbar$)
at a distance $d$ from the boundary. Thus, in the middle of the
condensate the correction term is of order $\exp(-1/\epsilon)$,
which is  nonanalytic in the small parameter
$\epsilon=\frac{1}{kR}=\sqrt{\frac{\pi R\hbar^2}{3 g m N}}$ 
($R$ being the radius of the spherical box).
It can be shown that one gets a similar nonanalytic boundary induced
correction in case of smooth trapping potentials, too.

Let us add that most of the above observations also apply
to the recently studied Bose condensates in optical traps
where the Zeeman energy is much smaller than the interaction energy.
Under such circumstances the total spin 
can freely rotate and new phenomena appear\cite{Ho}-\cite{sp5}. Among others,
for the total spin $f=1$ (as in case of $^{23}$Na and
$^{87}$Rb)
two types of ground states become possible.
These are called 'polar' and 'ferromagnetic' \cite{Ho}
and possess the form $\Phi_0 \zeta$ where $\zeta$ is a normalized 
three component spinor. These ground states correspond to the 
spontaneous breaking of $U(1)\times S^2$ and $SO(3)$ symmetry,
respectively.
For simplicity, we consider a zero temperature situation.
The substitute of the Gross-Pitaevskii equation is \cite{Ho}
\begin{eqnarray}
\left[-\frac{\hbar^2}{2m}\triangle + V(\vec{r})-\mu + b\Phi_0^2(\vec
{r})
\right]\Phi_0(\vec{r})=0
\label{GPp}
\end{eqnarray} 
where $b=c_0$
for the polar ground state and $b=c_0+c_2$ 
for the ferromagnetic ground state. Note that $c_0=(g_0+2g_2)/3$
and $c_2=(g_2-g_0)/3$
with $g_0=4\pi \hbar^2 a_0/m$ and $g_2=4\pi \hbar^2 a_2/m$,
where $a_0$ and $a_2$ are the $s$-wave scattering lengths 
in the total spin $F=0$ and $F=2$ channel, respectively.  
The polar ground state emerges if $c_2>0$, while for $c_2<0$
the ferromagnetic ground state becomes possible.
Collective modes can be found in both the polar and the 
ferromagnetic ground state as solutions 
of the corresponding generalized Bogolyubov equations.
They are the density modes in both the polar and ferromagnetic states
and the spin wave mode in the polar state. The mathematical structure
of the Bogolyubov equations is identical with the scalar case
discussed above, therefore, all our previous considerations
apply. In contrast, the equations, which are related to spin and
quadrupolar spin fluctuations in the ferromagnetic phase have
different structures. The modes have been found in the homogeneous
system \cite{Ho} and are determined in the presence
of the external potential by the equations
\begin{eqnarray}
\left[-\frac{\hbar^2}{2m}\triangle + V(\vec{r})-\mu + g_2\Phi_0^2(\vec
{r})
\right]u^0_j
=E_j u^0_j\label{Bof0}
\end{eqnarray}
and
\begin{eqnarray}
\left[-\frac{\hbar^2}{2m}\triangle + V(\vec{r})-\mu + (g_2+2|c_2|)\Phi_0^2(\vec
{r})
\right]u^{-}_j
=E_j u^{-}_j\label{Bofm1}
\end{eqnarray}
for the spin and quadrupolar spin modes, respectively. 

A combination 
of Eq.(\ref{Bof0}) with
the Gross-Pitaevskii equation (\ref{GPp}) shows that the bulk correction 
to the Thomas-Fermi approximation is important in this case, too. 
Taking this correction into account, we get for the
spin fluctuations $\delta M_-=\Phi_0^{TF} u_j^{0*}$
\begin{eqnarray}
-\frac{\hbar}{2m}\nabla \left(\left(\Phi_0^{TF}\right)^2\nabla 
\left(\frac{\delta M_-}{\left(\Phi_0^{TF}\right)^2}\right)\right)=
\omega_j \delta M_-\label{hyds}
\end{eqnarray}
This is similar to (but not identical with) the hydrodynamical equation \ref{hyd}.
It is interesting to note that $\Phi_0(\vec{r})$ itself is a solution of the eigenvalue equation
(\ref{Bof0}) with $E_j=0$ (cf. Eq.(\ref{GPp})) which corresponds to
the solution of Eq.(\ref{hyds}) $\delta
M_-=\left(\Phi_0^{TF}\right)^2$
with $\omega_j=0$.  
In case of 
quadrupolar
spin fluctuations Eq.(\ref{Bofm1}) is relevant.
Inserting Eq.(\ref{GPp}) the lowest order term $2|c_2|\left(\Phi_0^{TF}\right)^2$
does not cancel, therefore one would think
bulk corrections are negligible in this case, along with the kinetic
energy of the mode in the bulk. For an extremely
large condensate this is indeed the case for the lowest excitations. Nevertheless,
the numerical value of $c_2$ in case of $^{87}$Rb (which is the only
possible candidate for having a ferromagnetic ground state and $f=1$
hyperfine spin) is  
$ -(9 \pm 25)\times 10^{-3}g_2 $
, very small. Note that if 
$c_2$ is actually not negative, no ferromagnetic ground state 
is possible. 
The higher order terms (the kinetic term, for instance) can be
estimated by $\frac{\hbar^2}{2mL^2}=\frac{\hbar^2\omega_r^2}{4\mu}$, where $L$ is 
the major axis of the ellipsoidal
condensate. On the other hand, $2|c_2|\left(\Phi_0^{TF}\right)^2\approx 2|c_2|\mu/g_2$.
This will be larger than the other terms if
\begin{eqnarray}
\sqrt{8\frac{|c_2|}{g_2}}\approx 0.3 >\frac{\hbar\omega_r}{\mu}\;.\label{cond}
\end{eqnarray}
Here $\omega_r$ is the smallest trapping frequency. For $\omega_r=20 Hz$
we get $\hbar\omega_r/\mu=(10^4/N)^{2/5}$, thus for $N>2\times10^5$   
Eq.(\ref{cond}) is satisfied. 
For larger particle numbers bulk corrections do not play a role for
the lowest lying excitations. The kinetic term is also negligible, 
which looks rather paradoxical. The solution is 
that these excitations are confined to the vicinity of the
Thomas-Fermi surface as noticed already by Ho \cite{Ho}. For smaller
particle numbers bulk corrections must be taken into account, and when
the term $2|c_2|\Phi_0^2$ becomes negligible, the spectra of (\ref{Bof0})
and (\ref{Bofm1}) coincide.
Similarly to the case of spin fluctuations (cf.Eq.(\ref{hyds})) 
one can get the equation describing the quadrupolar spin
waves in the bulk of the condensate as follows:
\begin{eqnarray}
-\frac{\hbar}{2m}\nabla \left(\left(\Phi_0^{TF}\right)^2\nabla 
\left(\frac{\delta M^2_-}{\left(\Phi_0^{TF}\right)^2}\right)\right)\nonumber\\
+2\frac{|c_2|}{\hbar}\left(\Phi_0^{TF}\right)^2\delta M^2_-=
\omega_j \delta M^2_-\label{hydsq}
\end{eqnarray}
 
Here $\delta M^2_-=2\Phi_0^{TF}u_j^{-*}$. 
It is interesting to look at the relative magnitude of the terms in
the left hand side of Eq.(\ref{hydsq}). The second term is dominating
if Eq.(\ref{cond}) holds.
In any case its value in the lowest energy state (i.e. the gap for
excitations confined to the bulk) can be estimated by
using the Ansatz $u_j^{-0}=\frac{1}{\sqrt{N}}\Phi_0^{TF}$. One
obtains for the gap $\frac{2|c_2|}{N}\int
d^3\vec{r}\left(\Phi_0^{TF}\right)^4$. As the derivation
is independent of the form of the external potential, this result also applies in the
homogeneous case. Then we get back the known result 
$2|c_2||\Phi_0|^2$ for the gap\cite{Ho}. Note that for a fixed harmonic external potential
the gap increases with the particle number as $N^{2/5}$ (which is
proportional to the average density and also to
$\left|\Phi_0^{TF}(0)\right|^2$ in this case).

It is of worth noting that Eq.(\ref{cond}) is also the condition
for the applicability of the hydrodynamic approximation 
in case of the spin fluctuations in the polar state.
Due to the smallness of 
$c_2$ Eq.(\ref{cond}) is a more confining condition 
than the similar conditions in the scalar case
and in the case of density fluctuations in the polar  and
ferromagnetic states. Namely, the condition for the latter ones is similar to 
(\ref{cond}),
but $c_2$ should be replaced by $c_0$. This latter condition is also
relevant to the applicability of Eqs. (\ref{hyds}) and (\ref{hydsq}).


Fruitful discussions with R.Graham and M.Fliesser are gratefully
acknowledged.
 The  work has been partially supported by the 
Hungarian National Research Foundation under Grant Nos. \mbox{T 029752} and
T 029552. One of us (Gy.B.) would like to thank the Hungarian Academy of Sciences
for support.

\end{document}